\documentclass[a4paper]{article}

\usepackage{epsfig}
\usepackage{amssymb}

\begin{document}

\title{Instability of attractors in auto--associative networks with
bio--inspired fast synaptic noise}
\author{J.J. Torres\footnote{Journal Reference: LNCS 3512: 161-167, 2005. Corresponding author: Joaquin J. Torres, Department of Electromagnetism and Physics of the Matter, University of Granada, E-18071 Granada Spain; Tlf: (+34) 958244014; Fax: (+34) 95824 23 53 \hbox{email: jtorres@onsager.ugr.es}}, J.M. Cortes\footnote{Present address: Institute for Adaptive and Neural Computation, School of Informatics, The University of Edinburgh, 5 Forrest Hill, Edinburgh EH1 2QL, Scotland UK} and J. Marro\\
\small Institute \textit{Carlos I} for Theoretical and Computational
Physics, and \\
\small Departamento de Electromagnetismo y F\'{\i}sica de la Materia,\\
\small University of Granada, E-18071--Granada, Spain.}
\date{}
\maketitle

\begin{abstract}

We studied auto--associative networks in which synapses are \textit{noisy}
on a time scale much shorter that the one for the neuron dynamics. In our model
a presynaptic noise causes postsynaptic depression as recently observed in
neurobiological systems. This results in a nonequilibrium condition in which
the network sensitivity to an external stimulus is enhanced. In particular,
the fixed points are qualitatively modified, and the system may easily scape
from the attractors. As a result, in addition to pattern recognition, the
model is useful for class identification and categorization.
\end{abstract}

\section*{Introduction and model}

It is likely that the reported short--time \textit{synaptic noise}
determines the transmission of information in the brain \cite%
{abb,allenPNAS,zadorJN,bibitchkov}. By means of a modified attractor neural
network, we shall illustrate here that fast synaptic noise may result in a
nonequilibrium condition \cite{marroB} consistent with short--time
depression \cite{tsodyksNC}. We then show how this in turn induces escaping
of the system from the attractor. The fact that the stability of fixed
points is dramatically modified, in practice allows for complex
computational tasks such as class identification and categorization, in
close similarity to the situation reported in neurobiological systems \cite%
{monkey,animals,olfact}. A more detailed account of this work will be
published elsewhere \cite{FNseq0}.

Consider a set of $N$ binary neurons with configurations $\mathbf{S}\equiv
\{s_{i}=\pm 1;i=1,\ldots ,N\}$ connected by synapses of intensity 
\begin{equation}
w_{ij}=\overline{w}_{ij}x_{j}\,\,\,\,\forall i,j.  \label{ws}
\end{equation}%
Here, $\overline{w}_{ij}$ is fixed and determined in a previous \textit{\
learning} process, and $x_{j}$ is a stochastic variable. For fixed $\mathbf{%
W\equiv }\{\overline{w}_{ij}\},$ the network state at time $t$ is determined
by $\mathbf{A=(S,X}\equiv \{x_{i}\}).$ These evolve in time according to%
{\small 
\begin{equation}
\frac{\partial P_{t}(\mathbf{A})}{\partial t}=\sum_{\mathbf{A}^{\prime }}
\,\left[ P_{t}(\mathbf{A}^{\prime })c(\mathbf{A}%
^{\prime }\rightarrow \mathbf{A})-P_{t}(\mathbf{A})c(\mathbf{A}\rightarrow 
\mathbf{A}^{\prime })\right]   \label{gme}
\end{equation}%
where }$c(\mathbf{A}\rightarrow \mathbf{A}^{\prime })=p\,c^{\mathbf{X}}(%
\mathbf{S}\rightarrow \mathbf{{S^{\prime })\,}}\delta _{\mathbf{X,{X^{\prime
}}}}+(1-p)\,c^{\mathbf{S}}(\mathbf{X}\rightarrow \mathbf{{X^{\prime })}}%
\,\delta _{\mathbf{S},\mathbf{{S^{\prime }}}}$ \cite{torresPRA}. This
amounts to assume that neurons $(\mathbf{S})$ change stochastically in time
competing with a noisy dynamics of synapses $(\mathbf{X}),$ the latter with
an \textit{a priory} relative weight of $(1-p)/p$.

For $p=1,$ the model reduces to the Hopfield case, in which synapses are
quenched, i.e., $x_{i}$ is constant and independent of $i.$ We are
interested here in the limit $p\rightarrow 0$  for which neurons evolve as
in the presence of a steady distribution for the noise $\mathbf{X.}$ If we
write $P(\mathbf{S},\mathbf{X})=P(\mathbf{X}|\mathbf{S})\,P(\mathbf{S}),$
where $P(\mathbf{X}|\mathbf{S})$ stands for the conditional probability of $%
\mathbf{X}$ given $\mathbf{S,}$ one obtains from (\ref{gme}), after
rescaling time $tp\rightarrow t$ and summing over $\mathbf{X}$ that 
\begin{equation}
\frac{\partial P_{t}(\mathbf{S})}{\partial t}=\sum_{\mathbf{{S^{\prime }}}%
}\left\{ P_{t}(\mathbf{{S^{\prime }}})\bar{c}[\mathbf{{S^{\prime }}}%
\rightarrow \mathbf{S}]-P_{t}(\mathbf{S})\bar{c}[\mathbf{S}\rightarrow 
\mathbf{{S^{\prime }}}]\right\} .  \label{geme}
\end{equation}%
Here, $\bar{c}[\mathbf{S}\rightarrow \mathbf{{S^{\prime }}}]\equiv \sum_{%
\mathbf{X}}\,P^{\mathrm{st}}(\mathbf{X}|\mathbf{S})\,c^{\mathbf{X}}[\mathbf{S%
}\rightarrow \mathbf{{S^{\prime }}}],$and the stationary solution is%
\begin{equation}
P^{\mathrm{st}}(\mathbf{X}|\mathbf{S})=\frac{\sum_{\mathbf{X}}\,c^{\mathbf{S}%
}[\mathbf{{X^{\prime }}}\rightarrow \mathbf{X}]\,P^{\mathrm{st}}(\mathbf{{%
X^{\prime }}}|\mathbf{S})}{\sum_{\mathbf{X}}\,c^{\mathbf{S}}[\mathbf{X}%
\rightarrow \mathbf{{\ X^{\prime }}}]}.  \label{pst}
\end{equation}%
This involves an adiabatic elimination of fast variables; see technical
details in Ref.\cite{marroB}, for instance.

Notice that $\bar{c}[\mathbf{S}\rightarrow \mathbf{{S^{\prime }}}]$ is a
superposition. One may interpret that different underlying dynamics, each
associated to a different realization of the stochasticity $\mathbf{X,}$
compete. In the limit $p\rightarrow 0,$ an \textit{effective} rate results
from combining \textbf{$c^{\mathbf{X}}[\mathbf{S}\rightarrow \mathbf{\ {%
S^{\prime }}}]$ }with probability\textbf{\ $P^{\mathrm{st}}(\mathbf{X}|%
\mathbf{S})$} for varying $\mathbf{X.}$ Given that each elementary dynamics
tends to drive the system to a different equilibrium state, the results is,
in general, a nonequilibrium steady state \cite{marroB}. The question is if
such a competition between synaptic noise and neural activity is at the
origin of some of the computational strategies in neurobiological systems.

For simplicity, we shall consider here \textit{spin--flip} dynamics for the
neurons, namely, stochastic local inversions $s_{i}\rightarrow -s_{i}$ as
induced by a bath at temperature $T.$ The elementary rate then reduces $c^{%
\mathbf{X}}[\mathbf{S}\rightarrow \mathbf{{S^{\prime }}}]=\Psi \lbrack u^{%
\mathbf{\ X}}(\mathbf{S},i)],$ where we assume $\Psi (u)=\exp (-u)\Psi (-u),$
$\Psi (0)=1$ and $\Psi (\infty )=0$ \cite{marroB}. Here, $u^{\mathbf{X}}(%
\mathbf{S},i)\equiv 2T^{-1}s_{i}h_{i}^{\mathbf{X}}(\mathbf{S}),$ where $%
h_{i}^{\mathbf{X}}(\mathbf{S})=\sum_{j\neq i}\overline{w}_{ij}x_{j}s_{j}$ is
the net presynaptic current or local field on the (postsynaptic) neuron $i.$

Our interest here is in modeling \textit{noise} consistent with short-term
synaptic depression \cite{tsodyksNC,torresNC}. We therefore assume the noise
distribution $P^{\mathrm{st}}(\mathbf{X}|\mathbf{S})=\prod_{j}P(x_{j}|%
\mathbf{S})$ with 
\begin{equation}
P(x_{j}|\mathbf{S})=\zeta \left( \vec{\mathbf{m}}\right) \mathrm{\ }\delta
(x_{j}+\Phi )+\left[ 1-\zeta \left( \vec{\mathbf{m}}\right) \right] \mathrm{%
\ }\delta (x_{j}-1).  \label{genbimod}
\end{equation}%
Here, $\vec{\mathbf{m}}=\vec{\mathbf{m}}(\mathbf{S})\equiv \left( m^{1}(%
\mathbf{S}),\ldots ,m^{M}(\mathbf{S})\right) $ is the $M$-dimensional
overlap vector, and $\zeta \left( \vec{\mathbf{m}}\right) $ stands for a
function of $\vec{\mathbf{m}}$ to be determined. The depression effect here,
namely, $x_{j}=-\Phi ,$ depends on the overlap vector which measures the net
current arriving to postsynaptic neurons. Consequently, the non--local
choice (\ref{genbimod}) introduces non--trivial correlations between
synaptic noise and neural activity.

This new case also reduces to the Hopfield model but only in the limit $\Phi
\rightarrow -1$ for any $\zeta \left( \vec{\mathbf{m}}\right) .$ In general,
however, the competition results in a rather complex nonequilibrium
behavior. As far as $\Psi (u+v)=\Psi (u)\Psi (v)$ and $\mathbf{P^{\mathrm{st}%
}}\left( \mathbf{\mathbf{X}|\mathbf{S}}\right) $ factorizes as indicated,
time evolution proceeds by the effective transition rate%
\begin{equation}
{\bar{c}}[\mathbf{S}\rightarrow \mathbf{S}^{i}]=\exp \left( -s_{i}h_{i}^{%
\mathrm{eff}}/T\right) ,  \label{etr2}
\end{equation}%
where 
\begin{equation}
h_{i}^{\mathrm{eff}}=\sum_{j\neq i}w_{ij}^{\mathrm{eff}}s_{j}.
\label{gloelf}
\end{equation}%
Here, $w_{ij}^{\mathrm{eff}}=\left\{ 1-\frac{1+\Phi }{2}\left[ \zeta \left( 
\vec{\mathbf{m}}\right) +\zeta \left( \vec{\mathbf{m}}^{\mathbf{i}}\right) %
\right] \right\} \overline{w}_{ij},$ are the effective synaptic intensities
as modified by the noise; $\vec{\mathbf{m}}=\vec{\mathbf{m}}(\mathbf{S}),$ $%
\vec{\mathbf{m}}^{\mathbf{i}}\equiv \vec{\mathbf{m}}(\mathbf{S}^{\mathbf{i}%
})=\vec{\mathbf{m}}-2s_{i}\vec{\xi}_{i}/N,$ and $\vec{\xi}_{i}=\left( \xi
_{i}^{1},\xi _{i}^{2},...,\xi _{i}^{M}\right) ,$ $\mathbf{\xi }^{\nu }=\{\xi
_{i}^{\nu }=\pm 1,i=1,\ldots ,N\},$ stands for the binary $M$--dimensional
stored pattern. 

In order to obtain the effective fields (\ref{gloelf}), we linearized the
rate $\bar{c}[\mathbf{S}\rightarrow \mathbf{{S^{\prime }}}]$ around $%
\overline{w}_{ij}=0.$ This is a good approximation for the Hebbian
prescription $\overline{w}_{ij}=N^{-1}\sum_{\nu }\xi _{i}^{\nu }\xi
_{j}^{\nu }$ as far as this only stores completely uncorrelated, random
patterns and for a sufficiently large system, e.g., in the thermodynamic
limit $N\rightarrow \infty .$ To proceed further, we need to determine a
convenient function $\zeta $ in (\ref{genbimod}). In order to model
activity--dependent mechanisms acting on the synapses, $\zeta \mathbf{\ }$
should be an increasing function of the field. In fact, this simply needs to
depend on the overlaps. Furthermore, $\zeta \left( \vec{\mathbf{m}}\right) $
is a probability, and it needs to preserve the $\pm 1$ symmetry. A simple
choice is 
\begin{equation}
\zeta \left( \vec{\mathbf{m}}\right) =\frac{1}{1+\alpha }\sum_{\nu }\left[
m^{\nu }\left( \mathbf{S}\right) \right] ^{2},  \label{psi}
\end{equation}%
where $\alpha =M/N.$ We describe next the behavior that ensues from (\ref%
{gloelf})--(\ref{psi}) as implied by the noise distribution (\ref{genbimod}).

The effective rate (\ref{etr2}) may be used in computer simulations, and it
may also be substituted in the relevant equations. Consider, for instance,
the \textit{overlaps,} defined as the product of the current state with one
of the stored patterns, $m^{\nu }({\mathbf{S}})\equiv \frac{1}{N}%
\sum_{i}s_{i}\xi _{i}^{\nu }.$After using standard techniques, it follows
from (\ref{geme}) that%
\begin{equation}
\partial _{t}m^{\nu }=2N^{-1}\sum_{i}\xi _{i}^{\nu }\sinh \left( h_{i}^{%
\mathrm{eff}}/T\right) -s_{i}\cosh \left( h_{i}^{\mathrm{eff}}/T\right),
\label{kinetic}
\end{equation}%
which is to be averaged over both thermal noise and pattern realizations.

\section*{Discussion of some main results}

We here illustrate the case of a single stored pattern, $M=1.$ After using
the simplifying (mean-field) assumption $\langle s_{i}\rangle \approx s_{i},$
one obtains from (\ref{etr2})--(\ref{kinetic}) the steady overlap $m^{\nu
=1}\equiv m=\tanh \{T^{-1}m\left[ 1-\allowbreak (m)^{2}\left( 1+\Phi \right) %
\right] \}.\ $This depicts a transition from a \textit{ferromagnetic--like}
phase, i.e., solutions $m\neq 0,$ to a \textit{paramagnetic--like} phase, $%
m=0.$ The transition is continuous or second order only for $\Phi >\Phi
_{c}=-4/3,$ and it then follows a critical temperature $T_{c}=1$ \cite%
{FNseq0}.

It is to be remarked that a discontinuous phase transition allows for a much
better performance of the retrieval process than a continuous one. This is
because the behavior is sharp just below the transition temperature in the
former case. Consequently, the above indicates that our model performs
better for large negative $\Phi ,$ $\Phi <-4/3.$ These results are in full agreement 
with Monte Carlo simulations of neural networks with fast 
presynaptic noise and using asynchronous sequential updating.

We also investigated the sensitivity of the system under an external
stimulus. A high sensitivity will allow for a rapid adaptation of the
response to varying stimuli from the environment, which is an important
feature of neurobiological systems. A simple external input may be simulated
by adding to each local field a driving term $-\delta \xi _{i},\ \forall i,$
with $0<\delta \ll 1$ \cite{bibitchkov}. A negative drive for a single
pattern assures that the network activity may go from the attractor, $%
\mathbf{\xi ,}$ to its \textquotedblleft antipattern\textquotedblright , $-%
\mathbf{\xi }.$ 
\begin{figure}[t]
\centerline{\psfig{file=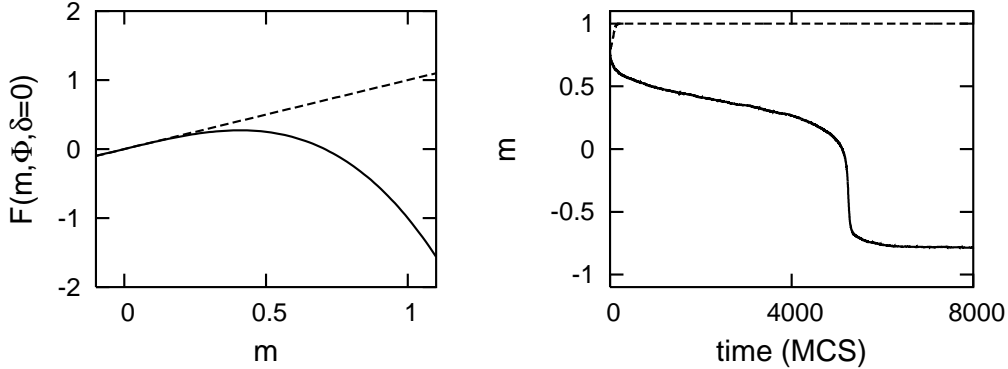,width=14cm}}
\caption{(Left) The function $F(m,\Phi ,\protect\delta =0),$ as defined in
the main text, for presynaptic noise with $\Phi =1$ (solid line) and in
absence of noise, i.e., $\Phi =-1$ (dashed line). Near the attractor ($%
m\approx 1$), $F$ is positive in the absence of noise, which leads
stability of the memory. However, the noise makes $F$ negative,
which induces instability of the attractor. (Right) A retrieval
experiment in a network of $N=3600$ neurons under external stimulation for $%
T=0.1,$ $\protect\delta =0.3,$ and the same values of $\Phi $ as in the left
graph.}
\label{fig2}
\end{figure}

It follows for $M=1$ the stationary overlap $m=\tanh [T^{-1}F(m,\Phi ,\delta
)]$ with $F(m,\Phi ,\delta )\equiv m[1-(m)^{2}(1+\Phi )-\delta ].$The left
graph of figure \ref{fig2} shows this function for $\delta =0$ and $\Phi =1$
(fast noise) and $\Phi =-1$ (Hopfield case). Depending on the sign of $F,$
there are two different types of behavior, namely, (local)stability $(F>0)$
and instability $(F<0)$ of the attractor, which corresponds to $m=1.$ That
is, the noise induces intrinsic instability resulting in switching between
the pattern and the antipattern when a small perturbation $\delta $ is
added. 

In general, adding the fast noise destabilizes the fixed point for the
interesting case of small $\delta$ far from criticality ($T\ll T_{c}$).
This is illustrated by Monte Carlo simulations of a network of $N=3600$
neurons, $M=1,$ $\delta =0.3$ and $T=0.1,$ as shown in figure \ref{fig2}
right. Starting from an initial condition near the attractor, the system
jumps to the antipattern when fast noise is present (solid line), and
remains in the attractor for the Hopfield case (dashed line).

\begin{figure}[t]
\vspace*{-0.5cm} 
\centerline{\psfig{file=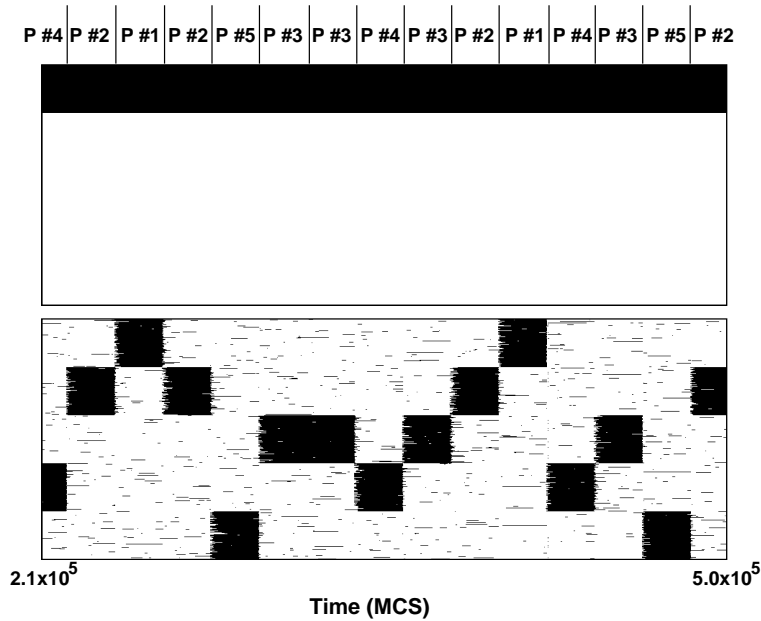,width=10cm}}
\caption{Sensitivity of the system under repetitive external random
stimulus, as discussed in the main text. The top graph shows the Hopfield
case ($\Phi =-1).$ Here, neuron activity is represented at vertical axe and simulation parameters are $N=400,$ $T=0.1$ and $\delta=0.3.$}
\label{fig3}
\end{figure}

The switching property remains as the system stores more patterns. In order
to illustrate this, we simulated a network of $N=400$ neurons with $M=5$
overlapping patterns such that $m^{\nu ,\mu }\equiv 1/N\sum_{i}\xi _{i}^{\nu
}\xi _{i}^{\mu }=1/5$ for any two of them. The system in this case begins
with the first pattern, and then evolves under the effect of a repetitive
small stimulus $+\delta \mathbf{\xi }^{\sigma }$ ($\delta =0.3$) with $%
\sigma $ randomly chosen from $1$ to $5$ every $2\times 10^{4}$ MCS for each 
$\sigma .$ As shown in figure \ref{fig3}, lacking the noise ($\Phi =-1$),
the system remains in the initial pattern. However, in the presence of some
noise ($\Phi =0.05$ in this simulation), there is continous jumping from one
attractor to the other, every time the new attractor is presented in the
stimulus. This property is robust with respect to the type of patterns
stored \cite{FNseq0}. 

Summing up, equations (\ref{geme})--(\ref{pst}) provide a rather general
framework to model activity--dependent processes. We here briefly reported
on some consequences of adapting this to a specific case. In particular, we
studied a case which describes neurobiologically--motivated fast noise, and
study how this affects the synapses of an auto--associative neural network
with a finite number of stored patterns. Assuming a noise distribution with
a global dependence on the activity, (\ref{genbimod}), one obtains
non--trivial local fields (\ref{gloelf}) which lead the system to an
intriguing emergent phenomenology. We studied this case both analytically
and by Monte Carlo simulations using Glauber, \textit{spin--flip} dynamics 
\cite{marroB}. We thus show that a tricritical point occurs. That is, one
has (in the limit $\alpha \rightarrow 0$) first and second order phase
transitions between a ferromagnetic--like, retrieval phase and a
paramagnetic--like, non--retrieval phase. The noise also happens to induce a
nonequilibrium condition which results in an important intensification of
the network sensitivity to external stimulation. We explicitly show that the
noise may turn unstable the \textit{attractor} or fixed point solution of
the retrieval process, and the system then seeks for another attractor. This
behavior improves the network ability to detect changing stimuli from the
environment. One may argue that the process of categorization in nature
might follow a similar strategy. That is, different attractors may
correspond to different objects, and a dynamics conveniently perturbed by
fast noise may keep visiting the attractors belonging to a class which is
characterized by a certain degree of correlation between its elements. A
similar mechanism seems at the basis of early olfactory processing of
insects \cite{olfact}, and instabilities of the same sort have been
described in the cortical activity of monkeys \cite{monkey} and other cases 
\cite{animals}. We are presently studying further variations of the model
above.

\section*{Acknowledgments}

We acknowledge financial support from MCyT and FEDER (project No.
BFM2001-2841 and \textit{Ram\'{o}n y Cajal} contract).
%
% ---- Bibliography ----
%

\end{document}